# When is Enough Enough? A Proposed Termination Point for the Number of Replicates in Computational Simulations

Eric T. Lofgren[a,b], Kellen Myers[c,d], and Nina H. Fefferman[b,c,e]

[a] Paul G. Allen School for Global Health, Washington State University, Pullman, WA, USA
[b] US NSF Center for Analysis and Prediction of Pandemic Expansion (APPEX), University of Tennessee, Knoxville, 37996, USA
[c] National Institute for Mathematical and Biological Synthesis (NIMBioS), University of Tennessee, Knoxville, 37996, USA
[d] Department of Mathematics, Tusculum University, Greenville, TN, 37745, USA
[e] Departments of Mathematics & Ecology and Evolutionary Biology, University of Tennessee, Knoxville, TN, 37996-3140, USA

Corresponding author: Nina H. Fefferman, US NSF Center for Analysis and Prediction of Pandemic Expansion (APPEX), University of Tennessee, Knoxville, 37996, USA; nfefferm@utk.edu

Abstract
Computational simulation provides a powerful toolkit for in silico experimentation. However, while the field has developed best practices for the design and implementation of such models, there remains ambiguity in discussions about how to understand and/or interpret their results due to their inherent ability to overwhelm traditional frequentist statistics by simply increasing the number of trials simulated. This fails the discipline in two ways: first, it leaves the community unsure of what constitutes a best practice for uniform understanding, and second, it potentially overburdens computational studies that burn clock cycles solely to ensure "enough runs to satisfy peers" without any theoretical underpinning for a definition of "enough". We propose a simple and straightforward standard for when to stop simulating additional trials, the Ω test, designed to be analogous to the function of traditional frequentist P-tests. Community adoption of a reasonable and uniform standard will permit more efficient computational experimentation and clearly communication/interpretation of the findings discovered in this way.

## Basic Statement of the Problem

Advances in both technological capability and algorithmic design have enabled a steadily increasing volume of simulation studies across all disciplines of scientific research over the past few decades. Complementing both the development of rigorous theory and empirical experimentation, *in silico* studies allow interrogation of systems that are either too difficult or costly to manipulate and/or too complicated to be analytically tractable. Simulation models are, at their core, computational experiments – closer to empirical trials than to analytic models. Elements of simulation studies are included, excluded, and manipulated by the same principles of experimental design that apply across empirical STEM fields[1, 2]. One fundamental aspect of valid experimental design, however, is critically different between simulation and real-world experimentation: statistical evaluation of the results[3]. Empiricists are trained from academic infancy that best practices for their work rely on creating a plan for statistical testing and interpretation as part of their experimental design before they begin their study: selecting appropriate statistical tests for significance, conducting power analyses, and checking that their study system will satisfy required assumptions (e.g., satisfying particular distributions, or independence, etc.). It is therefore natural that scientists reading computational experimental studies, realizing their empirical nature, should also expect the same types of statistical analyses to guide their interpretation of these simulated results[4]. In fact, journal standards and reviewer criticisms frequently require the inclusion of reported P-values or confidence intervals in order for a manuscript to be accepted for publication. Unfortunately, while well-intentioned, this is most often completely inappropriate for simulation studies (e.g. Lofgren[5], Hasley[6], Hasley et al.[7]).

Frequentist statistical tests are predicated on power analyses, estimating the ability of a sample to discriminate deviation from the null hypothesis based on the variance that should be represented in a randomly selected sample of the size considered[8]. Power is a function of the number of samples, effect size, the statistical test being used, and a pre-set value for α, the level of significance desired for the statistical test [9, 10]. Power increases with sample size. The same slight difference that would be analyzed as "not significant" with a small number of samples will be found to be significant by the same test, with the same threshold for significance, when the number of samples becomes sufficiently large. Frequently, sample size is also the only means at hand for an empirical researcher to increase the power of their study – and increasing sample size often requires a non-trivial increase in the cost and complexity of a study.

For a computational experiment, in contrast, for any fixed threshold of significance, finding that the data support a significant conclusion (i.e., rejection of the null hypothesis) is simply a matter of running the model a sufficient number of times[4]. While it's true that some studies would need to be run so many times that the computational costs/time exceeds a reasonable duration for experimentation (even on something like a high-performance computing cluster), those are very rare relative to the studies where a day or week of additional computational investment will guarantee a significant result, increasing the chances of having a publishable finding. It is therefore clear that methods depending on significance testing are inappropriate for analyzing and interpreting simulation/computational experiments.

Not all is lost; nonparametric and Bayesian methods can provide appropriate guidance for interpretation of computational results. Some studies have proposed simply reporting the power analysis associated with replicate number used[4], or estimating "accuracy in power estimations"[11], but these still leave many open-ended choices and allow for confusion in conformity of best practices across different types of studies. Ultimately, there remains one critical challenge: how to know when there is sufficient outcome data that the analyzed results reflect the impacts of the focal drivers of the system being studied and are not the result of the accumulation of (perhaps synergistic) stochastic noise in the variables, parameters, and interactions being simulated. In essence, how can we design a power analysis for appropriate tests for simulation studies that can overwhelm traditional methods by replication?

## How Many Replicates are Enough?

To be of greatest use, any designed threshold for determination of how many replicated runs of a simulation study are enough to provide reliable analysis and interpretation should be rigorous, generalizable, and quantifiable for any simulated system. It should be easy to define and evaluate by modelers and reviewers, and create a standard for best practice that can be used for interpretability across studies. Luckily, the same underlying logic that provides the foundation for frequentist power analyses can be applied here to suggest exactly such a threshold for simulation studies. Even for nonparametric studies, the most basic unit for understanding how to compare sets of outcomes is variance[12]: how much do we expect datapoints within a group to fluctuate in relation to each other and in comparison to datapoints from other groups to which they will be compared. Variance is no less meaningful when the datapoints are generated by computational simulation than they are in the real world, but the construct of a computational experiment allows far more nuanced consideration of how variance is understood. We can leverage that additional nuance.

In a computational simulation, it is easy to calculate the variance in the focal outcome variables with each calculated replicate of the experiment. Since each replicate is the production of one datapoint (i.e., sample), increasing the number of replicates will eventually lead to convergence in the estimated variance over all replicates. The only nuance is to ask how to rigorously characterize this convergence in a way that is generalizable across systems – we propose that using relative variance (i.e., normalizing the variance by the mean value of the variable) allows us to think of expected percent change in estimated variance resulting from each next computational replicate. We can then impose an arbitrary, but universally acceptable, threshold in that percent change in relative variance over an equally arbitrary, but universal number of replicates (to smooth over local noise in a consistent way), in a way directly analogous to using a P-value threshold of 0.05 – arbitrary, but still useful. It may be worth noting that this number will change drastically from study to study and will almost never be a "nice round number" (i.e., it is just as likely to then report having performed 623 realizations of a model as 10,000 realizations). Rather than being an arbitrary but aesthetically pleasing number that seems large enough that it is unlikely that the true dynamics remain undiscoverable (as is currently frequently the case for published studies' replicate choices), there will be a rigorous, quantitative rationale for the number. Of course, many studies consider multiple outcome variables and the variance in the different outcomes may not converge simultaneously. It is therefore appropriate to take the most conservative value (i.e., the last variable to converge) and apply it uniformly for estimation of all measured outcome variables of the system. (Note: It will be critical to consider the variance of each of the precise outcomes individually, e.g., if one is interested in the average difference in some set of paired data, it will not be appropriate to measure the variance of each element of the pairs. Rather, one must directly measure the variance of the difference itself.) So long as you are above the minimum convergence threshold for each variable being tested, you are assured not to have impacted the results of your analysis by number of replicates.

How to Do the Simplest Version in Easy Steps

Step 1: Run $x$ replicates of your simulation experiment, yielding all $Y$ focal outcome variables to be studied (Note: sample sizes so small that the central limit theorem cannot apply will lead to noise in the variance convergence estimate, so it is good practice to start with $x \geq 30$.)

Step 2: For each $y^j \in Y$, calculate the mean value of the outcome variable $\underline{y^j} = \left| \frac{\sum_{i=1}^{x} y_i^j}{x} \right|$ and the variance $Var(y^j) = \frac{\sum_{i=1}^{x} \left(y_i^j - \underline{y^j}\right)^2}{x-1}$.

Step 3: For each $y^j \in Y$, calculate the percent relative variance at x replicates for the outcome variable $y^j$ by the sample mean: $Z_x^j = \frac{Var(y^j)}{\underline{y^j}} * 100$

Step 4: Run one more replicate, so the total is now $x$+1.

Step 5: Calculate the difference in $Z$ achieved by going from $x$ to $x$+1: $\Delta Z_{x,x+1}^j = \left| Z_x^j - Z_{x+1}^j \right|$

Step 6: For a fixed number of replicates Q, when it is true for all j that each of the elements in the series $\Delta Z_{x,x+1}^j, \ldots \Delta Z_{x+Q-1,x+Q}^j$ are each less than the threshold value $\Omega$, terminate the simulation

Critical Difference from Seemingly Similar Frequentist Pitfalls

While there are appealing analogies to the use of P-values and significance testing, there are a number of subtleties that may seem inappropriate or unintuitive if we cleave too closely to heuristics developed for frequentist simulation. The first, and perhaps most critical, is that you must continuously calculate the estimated variance of your outcomes in order to know when the simulation may be safely halted without a high likelihood of lingering, as-yet uncaptured dynamics. At first glance, this may seem very similar to the practice of "p-hacking" - performing a statistical analysis repeatedly with deliberate perturbations in hopes of finding a significant result[13]. It is important to consider these thresholds not as *a priori* specified standards, but instead as measurement instruments connected to the experiment, reporting back when your simulated data has converged – before which it is inappropriate to analyze and after which additional simulation runs will yield no further insight.

Once this threshold has been reached, it is almost always most appropriate to approach the analysis of simulated data using nonparametric statistics. The reason for this is that the primary concern that argues against the use of nonparametric statistics, their lower statistical power, is no longer nearly as relevant. Simultaneously, protection against the assumptions of parametric statistics and the absence of any guarantee that the distribution of results from a simulation study are at all "well behaved" in any statistical sense, become much more important. Indeed, there is an argument to be made that, having simulated the distribution of your outcome data such that the variance of that outcome data is stable, we are beginning to converge on a setting in which we have fully sampled from the relevant source distributions and statistical tests are no longer necessary at all. However, the authors recognize that this may seem like a step too far for many, hence recommending nonparametric methods as the default best practice.

## Recognizing the Unique Nature of Computational Experiments

In addition to the methodological changes that necessarily flow from the nature of computational experiments and the unsuitability of P-value-dependent hypothesis testing, there are also necessary cultural changes. Paper reviewers and journal editors, especially those who are not simulation experts themselves, often request P-values for these experiments or power calculations in manuscripts and proposals. These types of requests need pushback, both because these figures do not add information beyond evidence that the researcher has been sufficiently patient and, if conflated with the figures that arise from empirical studies, could give a false sense of assurance that the study has been properly designed and has fully explored the dynamics of the simulated system. Instead, when planning a computational experiment, it is appropriate to specify the *a priori* threshold that is considered appropriate for asserting the variance of the system has converged ($\Omega$), and when reporting results, both this threshold and the ensuing number of iterations. For experiments with many outcome variables, or for which it is particularly difficult to achieve convergence, one might also consider plotting the estimated variance by the number of simulation runs.

Our arguments parallel earlier work from the empirical literature, e.g., work on the occasional anti-conservative nature of a significance threshold of 0.05[14] or the "Table 2 Fallacy"[15]. It is our goal to complement earlier perspectives (e.g.,[16] ), to provide a citable reference to strengthen these arguments, and create a common definition about consistent best practices for simulation modeling.

## Conclusion

Computational experiments are a powerful addition to the tools available for research, but they also straddle the ground between theoretical and empirical studies in ways that add complexity. One of the most pervasive difficulties in application of these tools is in their interpretability. By proposing our $\Omega$ test, we hope to decrease ambiguity and increase adoption of uniform best practices for interpretability, critically reducing both computational resources needed to accomplish research goals and, more importantly, intellectual strain and the potential for errant conclusions on researchers, reviewers, and anyone looking to gain / apply insight from the fruits of simulation methods.

Funding:

This work was supported by the National Science Foundation [DBI 2412115] to the NSF Center for Analysis and Prediction of Pandemic Expansion, the National Institutes of Health [R35GM147013], and the National Institute for Mathematical and Biological Synthesis (which was founded with support by the National Science Foundation [DBI 1300426]).